\documentclass[aps,prb,twocolumn,showpacs]{revtex4}
\usepackage{amsmath,bm}
\usepackage{graphicx,epsfig,braket,amssymb,amsfonts}
\DeclareGraphicsExtensions{.pdf,.png,.jpg}
\pdfoutput=1

\renewcommand{\v} [1]{{\bf #1}}
\newcommand{\ba}{\begin{eqnarray}}
\newcommand{\ea}{\end{eqnarray}}
\newcommand{\nn}{\nonumber \\}
\newcommand{\bpm}{\begin{pmatrix}}
\newcommand{\epm}{\end{pmatrix}}

\begin{document}

\title{Zero-energy Bound States in Nodal Topological Lattice}

\author{Soo-Yong Lee}
\affiliation{Department of Physics, Sungkyunkwan University, Suwon 440-746, Korea}
\author{Jung Hoon Han}
\email[Electronic address:$~~$]{hanjh@skku.edu}
\affiliation{Department of Physics, Sungkyunkwan University, Suwon 440-746, Korea}

\begin{abstract} Nodal topological lattice is a form of magnetic crystal with topologically non-trivial spin texture, which further exhibits a periodic array of nodes with vanishing magnetization. Electronic structure for conduction electrons strongly Hund-coupled to such nodal topological lattice is examined. Our analysis shows that each node attracts two localized states which form narrow bands through inter-node hybridization within the mid-gap region. Nodal bands carry a Chern number under suitable perturbations, suggesting their potential role in the topological Hall effect. Enhancement of the density of states near zero energy observable in a tunneling experiment will provide a signature of the formation of nodal topological lattice.
\end{abstract}
\maketitle

\section{Introduction}
Skyrmion crystal phase has been identified in chiral magnets minimally characterized by the Heisenberg ferromagnetic exchange and Dzyaloshinskii-Moriya interactions~\cite{Boni,Tokura,NagaosaTokura}. Magnetic field plays a key role for these materials inducing a transition from the spiral ground state to the triple spiral phase, equivalent to the triangular lattice of Skyrmions. Although this is the most widely discovered form of topological spin lattice in chiral magnets so far, theory suggests many other forms of possible topological lattices~\cite{Bogdanov,Vishwanath,Han1,Han2,Heinze}.  In two dimensions, superposition of two orthogonally propagating spirals results in the meron-anti-meron (M$\overline{\rm M}$) lattice where nodes (points of vanishing magnetization) form a periodic array~\cite{Han1,Han2,Heinze}. In three dimensions, multiple spiral phases are equivalent to a lattice of hedgehogs and anti-hedgehogs (H$\overline{\rm H}$) realizing simple cubic, or other crystal symmetries~\cite{Vishwanath,Han2}. Magnetization nodes accompany all of the known three-dimensional topological lattices constructed so far and remain robust against application of magnetic field, while the two-dimensional nodes are lifted by it. A recent observation of the Hall effect in MnGe is believed to be a consequence of two-dimensional (or three-dimensional) square (simple cubic) structure of M$\overline{\rm M}$ (H$\overline{\rm H}$) lattice formed in that material~\cite{Kanazawa,Tokura2,Mirebeau}.
\\

\begin{figure}[tb]
\includegraphics[width=0.47\textwidth]{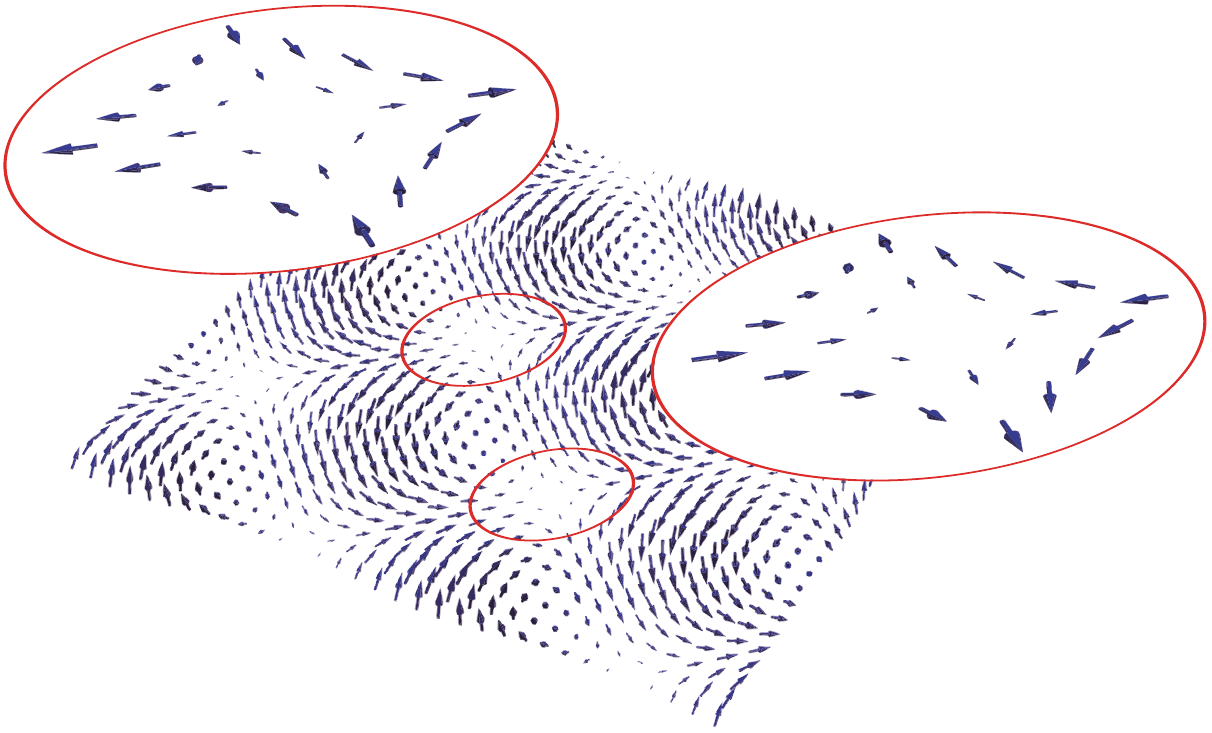}
\caption{(color online) Spin configuration of 2D M$\overline{\rm M}$ lattice. There are two nodal points in the unit cell. Spin structure around each node is an anti-vortex as shown.}\label{fig1}\end{figure}

Dynamics of conducting electrons coupled through strong Hund's rule exchange to the topological lattice of local moments displays a fascinating array of phenomena dubbed ``emergent electrodynamics"~\cite{NagaosaTokura}. Several of its stark predictions such as the topological Hall effect and the analogue of Faraday's law of induction have been confirmed experimentally~\cite{NagaosaTokura,Pfleiderer}. Despite its remarkable success, the theory of emergent electrodynamics suffers from a subtlety concerning the interaction of electrons with spins at the nodal points. Existing theories~\cite{JYe,Vishwanath1,Zang} rely on the large
Hund's exchange to kinetic energy ratio in assuming perfect alignment of local and itinerant spin moments. Such assumption obviously fails when the effective Hund coupling, given by the combined value $J |\v S_{\v r}|$, $J$=exchange energy, $\v S_{\v r}$=local magnetization, turns to zero at the node. How to get around the adiabatic assumption for the nodal region has remained unclear. In this paper we provide a fresh perspective on this issue by working out the electronic structure in interaction with the topological crystal phase with nodes, which we call the nodal topological lattice.
\\

The paper is organized as follows. In Sec. \ref{sec2} we outline the model for the nodal topological lattices. Localized zero energy states are shown numerically to exist in the vicinity of the nodal points of the 2D M$\overline{\rm M}$ and 3D H$\overline{\rm H}$ lattices. In Sec. \ref{sec3} we investigate the wavefunction of the zero energy states in the lattice model. Zero energy states are constructed explicitly for small $t/J$ on the lattice. In Sec. \ref{sec5} we show the full electronic structure of 2D M$\overline{\rm M}$ and 3D H$\overline{\rm H}$ lattices and discuss Chern numbers of the mid-gap bands. Finally, we discuss the experimental realization and application of the zero energy states and conclude in Sec. \ref{sec6}.

\section{Model for Nodal topological lattice}
\label{sec2}
The interplay of conduction electrons through Hund's coupling to the local moments is captured by a simple Hamiltonian

\ba H& = & H_K + H_J , \nn
H_K &=& -t \sum_{\langle \v{r},\v{r}'\rangle,\sigma=\uparrow,\downarrow} c^\dagger_{\v{r},\sigma} c_{\v{r}',\sigma} ,  \nn
H_J & = & - J \sum_{\v{r},\sigma,\sigma'} \v{S}_{\v{r}} \cdot \bigl( c^\dagger_{\v{r},\sigma} \bm{\sigma}_{\sigma,\sigma'} c_{\v{r},\sigma'} \bigr) . \label{eq:1}\ea
To the zeroth order in $t/J$ electronic energies occur at $\pm J|\v S_{ \v r}|$, leaving a gap between $+J|\v S_{\v r }|_{min.}$ and $-J|\v S_{\v r}|_{min.}$. The adiabatic theory of emergent electrodynamics rests on a sufficiently large value of $J|\v S_{\v r}|_{min.} /t$~\cite{JYe,Vishwanath1,Zang}. When there are nodes $|\v S_{\v r}|_{min.}=0$ there will be two zero-energy states per node in the limit of vanishing hybridization $t=0$, one for each spin orientation. As $t$ increases, these nodal states remain exactly at (2D), or quite near (3D), zero energy eventually forming bandgap.

\begin{figure}[tb]
\includegraphics[width=0.48\textwidth]{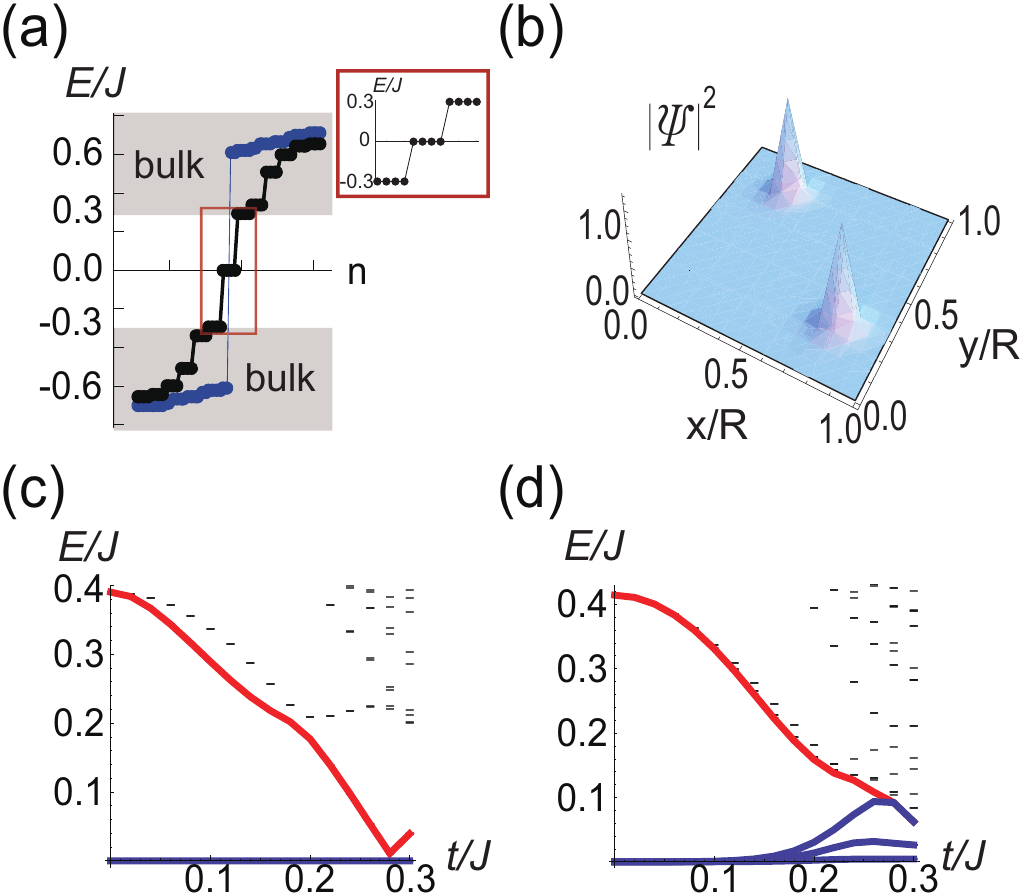}
\caption{(color online) (a) Energies of the 2D M$\overline{{\rm M}}$ lattice for $t/J=0.1$ and $R=16$ on a $R\times R$ lattice with periodic boundary conditions (black). $n$ is the order of the eigenenergies. There are four zero-energy states away from the bulk bands (Inset). There is no zero energy state in the case of single spiral, $\v{S}_\v{r}=(0, \cos Kx, \sin Kx)$ (blue). (b) Total electronic densities of the four localized zero energy solutions for $t/J=0.1$ and $R=16$. Dependence of energy spectra $E/J$ on $t/J$ in (c) 2D M$\overline{{\rm M}}$ lattice and in (d) 3D H$\overline{{\rm H}}$ lattice. Calculations are done for $R \times R $ ($R \times R \times R$) lattice, $R=16$ ($R=8$), and under periodic boundary condition. Bottom of the bulk energy band shown as red curves are clearly separated from the localized states also colored for small $t/J$. There are 4 degenerate (16 nearly degenerate) localized states for the 2D M$\overline{{\rm M}}$ (3D H$\overline{{\rm H}}$) lattice.}\label{fig2}\end{figure}

Existence of nodal zero-energy states for finite $t/J$ can be confirmed by numerical diagonalization of the  model Hamiltonian, first focusing on the 2D M$\overline{{\rm M}}$ lattice~\cite{Han1}

\ba \v{S}^{{\rm M}\overline{{\rm M}}}_{\v{r}}= \Bigr( \sin K y, \cos K x, \sin K x+ \cos K y  \Bigl).\label{eq:2}\ea
Coordinates $\v{r}=(x,y)$ are integer-valued and $K=2\pi/R$ for some integer choice of Skyrmion radius $R$. Nodes appear at $\v r_1 = (1/4, 1/2)R$ and $\v r_2 = (3/4,0)R$ within the magnetic unit cell of size $R\times R$, both corresponding to local anti-vortex configurations in the immediate vicinity of nodes as shown in Fig. \ref{fig1} (one spin-up, one spin-down). Separation between the nodes is half the magnetic unit cell: $\bm \pi_2 = (1/2,1/2)R$. Choosing the lattice size equal to $R\times R$ and imposing periodic boundary condition, numerical solution of the model (\ref{eq:1})  containing one pair of meron and anti-meron structure is shown in Fig. \ref{fig2}. There are exactly four degenerate zero energy states inside the gap as shown in Fig. \ref{fig2} (a), each node attracting two localized solutions as shown in Fig. \ref{fig2} (b)~\cite{comment}.
On increasing $t/J$, the minimum energy of the bulk band decreases as shown in Fig. \ref{fig2} (c), but the zero energy states still remain exactly at $E=0$.

For the 3D H$\overline{{\rm H}}$ lattice of simple cubic symmetry given by~\cite{Han1}

\ba \v{S}^{{\rm H}\overline{{\rm H}}}_{\v{r}}= \v{S}^{{\rm M}\overline{{\rm M}}}_{\v{r}}+\Bigr( \cos K z, \sin K z,0 \Bigl) , \label{eq:3}\ea
nodes appear at eight points in the unit cell of size $R^3$. Hedgehogs are centered at $(1, 3, 5)R/8$, $(3, 5, 1)R/8$, $(5, 1, 3)R/8$, $(7, 7, 7)R/8$, while anti-hedgehog locations are displaced by $\bm \pi_3 = (1/2,1/2,1/2)R$ from each hedgehog position. Electronic spectra show a total of 16 localized solutions per unit cell (two per node). Energies of the localized states are shown in Fig. \ref{fig2} (d). The states are not exactly zero, but almost zero and well away from the bulk band in the small $t/J$ region. On increasing $t/J$, the minimum energy of the bulk band decreases and eventually connects with those of some of the localized states. Even when this happens, some of the localized solutions remain well-separated from the bulk spectra.
\\

We examined the stability of the zero energy state in the nodal 2D lattice. Due to thermal fluctuation or irregular magnetic order, the spin vectors can be slightly modified from the perfect form in Eq. (\ref{eq:2}) or Eq. (\ref{eq:3}). We randomly varied the spin amplitude by $\pm 10$ \% and spin angle in the range of $0.2 \pi$ from the original spin vectors, while the nodal position is held fixed. Due to the non-vanishing spin sum in the vicinity of the node (See the argument about the necessary condition for the zero energy state in Sec. \ref{sec3B}), all zero energy states slightly are lifted from exact zero. However the energy scale of the broadness due to the random fluctuation, $J \delta \left(|\v S_{\v r }|_{min.}\right)$ is much smaller than the gap, $J|\v S_{\v r }|_{min.}$ due to the Hund coupling in the small $t/J$.

\section{Zero-energy electronic states in nodal topological lattice}

\label{sec3}

\subsection{Symmetries of nodal topological lattice}
\label{sec3A}
Applying the time-reversal operation $\Theta \equiv -i \sigma_y \cal{K}$ (${\cal K}$=complex conjugation) on the Hamiltonian  (\ref{eq:1}) gives

\ba \Theta H_K = H_K \Theta, ~~ \Theta H_J = -H_J \Theta. \ea
The translation operator ${\cal{T}}$ that moves each point by half the topological lattice spacing $\v r \rightarrow \v r + {\bm \pi}_d$ ($d=2,3$) has the property
\ba {\cal{T}} H_K  = H_K {\cal{T}}, ~~ {\cal{T}} H_J = -H_J {\cal{T}} , \ea
following from the fact that $\v S_{\v r \!+\! {\bm \pi}_d } = -\v S_{\v r}$ for the spin structure under consideration. Taking the product of the two mutually commuting operators, $\Theta$ and $\cal T$ to form a quasi-Kramers operator $\Lambda= \Theta {\cal T}$ ($\Lambda^2=-{\cal T}^2$), one obtains $[H,\Lambda]=0$.  Further,
on a $R^d$ lattice with periodic boundary condition we also have ${\cal T}^2 = 1$, $\Lambda^2 =-1$, thus the two eigenstates of energy $E_n$, $|n\rangle$ and $\Lambda|n\rangle$ will indeed be Kramers pairs. Another symmetry operation ${\cal S}$, $c_{\v{r},\sigma} \rightarrow (-1)^{N_\v{r}} c_{\v{r},\sigma}$, has the property

\ba {\cal{S}} H_K = -H_K {\cal{S}}, ~~ {\cal{S}} H_J = H_J {\cal{S}} ,\ea
where $N_{\v r}$ is the minimal number of steps from the node. The composite operator $\Omega = \Theta {\cal S}$ ($\Omega^2 = -1$) anti-commutes with the Hamiltonian $ \{ H,\Omega \}=0$ while commuting to $\Lambda$, and serves as the particle-hole conjugation operator relating $|n\rangle$ with $\Omega| n\rangle$ of opposite  energy. Should a zero energy state $|\zeta\rangle$ exist, it must do so in quartet, $( |\zeta\rangle, \Lambda|\zeta\rangle, \Omega|\zeta\rangle, \Lambda\Omega|\zeta\rangle)$, in agreement with the numerical findings for two-dimensional M$\overline{\rm M}$ lattice for $R \times R$ lattice. In 3D H$\overline{\rm H}$ lattice, two of all states are always degenerate due to $\Lambda$ operation. The particle-hole symmetry also gives rise to the negative energy states, as the count-partners of the positive energy states.
\\

\subsection{ Construction of zero-energy states}
\label{sec3B}
Zero-energy solutions around a single node can be constructed  for small $t/J$. With the nodal position at the origin, the state has the expansion
\ba |\zeta \rangle \!=\! |\bm 0 , \v{S}_0 \rangle \!+\!  \sum_{\v{r}} \left(\frac{t}{J} \right)^{N_{\v r}}  \Bigl[c^+_{\v{r}}  |\v{r},\v{S}_{\v{r}} \rangle \!+\! c^-_{\v{r}} |\v{r},-\v{S}_{\v{r}} \rangle \Bigr] \label{eq:zeta-1}\ea
in terms of coherent states $|\v r, \pm \v S_{\v r}\rangle$ of either spin orientations $\pm \v S_{\v r}$ at site $\v r$. For the magnetization $\v S_\v{r}=|\v S_\v{r}|\left(\cos \phi_\v{r} \sin \theta_\v{r} , \sin \phi_\v{r} \sin \theta_\v{r}, \cos \theta_\v{r}\right)$, spin coherent states are given as
\ba |\v S_\v{r}\rangle&=&\left(
                           \begin{array}{c}
                             \cos \frac{\theta_\v{r}}{2} \\
                             e^{i \phi_\v{r}} \sin \frac{\theta_\v{r}}{2} \\
                           \end{array}
                         \right), ~ |-\v S_\v{r} \rangle = \left(
                                                               \begin{array}{c}
                                                                 -e^{-i \phi_\v{r}} \sin \frac{\theta_\v{r}}{2} \\
                                                                 \cos \frac{\theta_\v{r}}{2} \\
                                                               \end{array}
                                                             \right).~~~~\ea
Absence of spin at the node allows certain freedom in choosing the nodal spin orientation $\v S_0$. $N_{\v r}$ is the minimal number of steps in a path required to reach a given point $\v r$ starting from the node $\bm 0$ (Recall that all such paths have the same $N_{\v r}$). We have

\begin{widetext}
\ba H_K |\zeta \rangle &=&-t \sum_{\v{r},\v{e}_\alpha} \Biggl[ \left(\frac{t}{J} \right)^{N_\v{r}-1} \Bigl( c^+_{\v{r}-\v{e}_\alpha}\langle \v{S}_\v{r} | \v{S}_{\v{r}-\v{e}_\alpha} \rangle | \v{r}, \v{S}_\v{r} \rangle + c^-_{\v{r}-\v{e}_\alpha}\langle -\v{S}_\v{r} | -\v{S}_{\v{r}-\v{e}_\alpha} \rangle | \v{r}, -\v{S}_\v{r} \rangle \Bigr) , \nn
&&~~~~~~+  \left(\frac{t}{J} \right)^{N_\v{r}+1} \Bigl( c^+_{\v{r}+\v{e}_\alpha}  \langle \v{S}_\v{r} | \v{S}_{\v{r}+\v{e}_\alpha} \rangle | \v{r}, \v{S}_\v{r} \rangle + c^-_{\v{r}+\v{e}_\alpha}\langle -\v{S}_\v{r}| -\v{S}_{\v{r}+\v{e}_\alpha} \rangle | \v{r}, -\v{S}_\v{r} \rangle\Bigr) \Biggr],\label{eq:8} \\
H_J |\zeta \rangle &=&-J \sum_{\v{r} \neq 0}  \left(\frac{t}{J} \right)^{N_\v{r}}|\v{S}_\v{r}| \left[c^+_\v{r}| \v{r}, \v{S}_{\v{r}} \rangle-c^-_\v{r}| \v{r}, -\v{S}_{\v{r}} \rangle \right] ,\label{eq:9}
\ea
\end{widetext}
where $\v{r}-\v{e}_\alpha$ is the nearest-neighbor of $\v{r}$ lying one step closer to the origin than $\v{r}$, carrying the prefactor $( t/J )^{N_{\v{r}}-1}$, and $\v{r}+\v{e}_\alpha$ is one step further away from the origin with the factor $\left(t/J \right)^{N_{\v{r}}+1}$. $|\v{S}_{\v{r}}|$ is the magnetization amplitude at $\v{r}$. When deriving Eq. (\ref{eq:8}) we ignored $\langle -\v S_{\v r \pm \v{e}_\alpha } | \v S_{\v r} \rangle$ as small in magnitude compared to $\langle \v S_{\v r \pm \v{e}_\alpha} | \v S_{\v r}\rangle$, assuming a smoothly varying spin background. In order that the sum of Eqs. (\ref{eq:8}) and (\ref{eq:9}) vanish, $(H_K + H_J ) |\zeta \rangle = 0$, one must require the coefficient for each $|\v{r},\pm\v{S}_\v{r} \rangle$ vanish, i.e.,

\ba  && c^\pm_{\v{r}-\v{e}_\alpha} \langle \pm\v{S}_\v{r} | \pm\v{S}_{\v{r}-\v{e}_\alpha} \rangle+\left(\frac{t}{J} \right)^{2} c^\pm_{\v{r}+\v{e}_\alpha} \langle \pm\v{S}_\v{r} | \pm\v{S}_{\v{r}+\v{e}_\alpha} \rangle \nn
&& ~~~~~~~~~~~~~~~~~~~~~~~ =\mp c^\pm_\v{r} |\v{S}_\v{r}| ~~~~~~~ (\v r \neq \v 0 ) , \label{eq:1.13a}\\
&& \sum_{\v{e}_\alpha } \Bigl( c^+_{\v e_\alpha} | \v{S}_{\v e_\alpha} \rangle+c^-_{\v{e}_\alpha}|-\v{S}_{\v{e}_\alpha} \rangle \Bigr) =0 ~~ (\v r = \v 0 ) .  \label{eq:1.13b}\ea
The second term in Eq. (\ref{eq:1.13a}), of order $(t/J)^2$ smaller than the first, can be ignored for small $t/J$, leading to the final simple recursion relation

\ba c^\pm_{\v r} = ( \mp 1 ) \sum_{\v{e}_\alpha} c^\pm_{\v{r}-\v{e}_\alpha} \frac{\langle \pm \v{S}_{{\v{r}}} | \pm \v{S}_{\v{r}-\v{e}_\alpha} \rangle}{|\v{S}_{\v{r}}|} .\ea
Again the sum on the right extends only over those neighbors of $\v r$ that lie one step closer to the origin.
Evaluating $c^\pm_{\v{r}}$ depends on the knowledge of $c^\pm_{\v{r}-\v{e}_\alpha}$, at neighbors lying one step closer to the origin. Extending the relation all the way to the origin gives a discrete path sum,

\ba c^\pm_{\v r}=(\mp1)^{N_{\v r}} \sum_{\cal{C}} \frac{\langle \v \pm \v S_{\v r} | \pm \v S_{\v r ' } \rangle}{|\v{S}_{\v{r}}|} \cdots \frac{\langle \pm \v S_1 | \v S_0 \rangle}{|\v{S}_1|} . \label{eq:c}\ea
Each product inside the sum takes place along a particular path ${\cal C} = (\bm 0 \rightarrow \v r_1 \rightarrow \cdots \rightarrow \v r' \rightarrow \v r)$ from the node to $\v r$ of length $N_{\v r}$. Contributions of all equi-distant, distinct paths are then summed over to give $c^\pm_{\v r}$ above. Inserting $c^\pm_{\v e_\alpha} = \mp \langle \pm  \v S_{\v e_\alpha} | \v S_{\v 0} \rangle / |\v S_{\v e_\alpha} |$ for each nearest neighbour of the origin into Eq. (\ref{eq:1.13b}) gives the extra condition on the zero-energy state,

\ba \sum_{\v e_\alpha} \v S_{\v e_\alpha}/|\v S_{\v e_\alpha}| = 0 .  \label{eq:sum_S} \ea
The sum of nearest-neighbor spin directions must be zero, in order to guarantee the existence of zero-energy mode.

A second zero-energy solution is found by the particle-hole symmetry operation,

\ba
\Omega |\zeta \rangle\!&=&\! |\bm 0 , -\v{S}_0 \rangle \nn &+&  \sum_{\v{r}} \left(-\frac{t}{J} \right)^{N_{\v r}}  \Bigl[ \left(c^+_{\v{r}} \right)^* |\v{r},-\v{S}_{\v{r}} \rangle \!-\! \left(c^-_{\v{r}} \right)^* |\v{r},\v{S}_{\v{r}} \rangle \Bigr].
\ea

Examining Eq. (\ref{eq:c}) $c^\pm_{\v r}$ in light of the identity $( \langle \v S_{\v r'} | \v S_{\v r}\rangle )^* = \langle -\v S_{\v r'} | -\v S_{\v r} \rangle$, we conclude that $|c^+_{\v r} | = |c^-_{\v r}|$ at all points $\v r$ provided $\langle \v S_{{\v e}_\alpha} | \v S_0 \rangle$ is chosen to have the same amplitude as $\langle -\v S_{{\v e}_\alpha} | \v S_0 \rangle$ for all immediate neighbors ${\v e}_\alpha$ of the origin. While this is not a necessary condition to guarantee a zero-energy state, such choice of $\v S_0$, if possible, ensures $|c^+_{\v r} | = |c^-_{\v r}|$ for all $\v r$. The resulting zero-energy solution $|\zeta\rangle$ (as well as $\Omega|\zeta \rangle$) then would have the nice property that its spin average remains strictly orthogonal to the local magnetization:
\ba\v S_{\v r} \cdot \langle \zeta | \bm \sigma_{\v r} |\zeta \rangle = 0 . \ea
The zero-energy state, unable to favor either parallel or anti-parallel spin orientation, chooses to develop the polarization in the plane perpendicular to the local moment. In two dimension such choice of $\v S_0$ is possible provided the four neighboring spins all lie in the plane, leaving an orthogonal direction for $\v S_0$. In 3D H$\overline{\rm H}$ lattice choosing such $\v S_0$ is generally impossible although zero-energy states can still be constructed quite easily with Eq. (\ref{eq:sum_S}).

\begin{figure}[tb]
\includegraphics[width=0.40\textwidth]{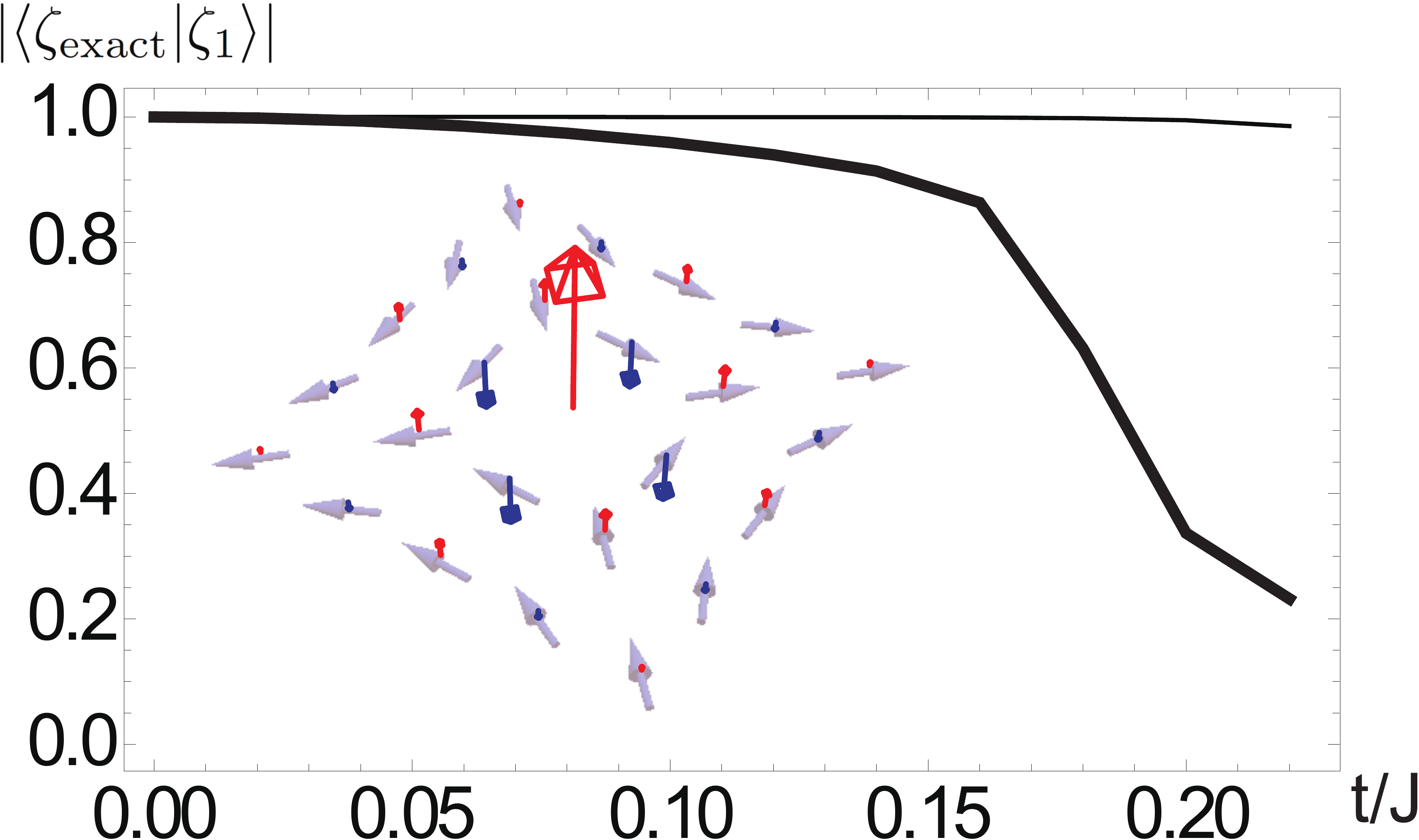}
\caption{(color online) Wavefunction overlap between the normalized zero energy state by the exact numerical result of lattice model and our analytic solution in Eq (\ref{eq:zeta-1}), $|\langle \zeta_{\mathrm{exact}}|\zeta\rangle|$. In the small $t/J$ limit, two zero energy state are almost identical. The calculation is performed under the spin texture when $\v{S}_{\v{r}}= f(r) \left(y/r, x/r,0 \right)$ in 2D (solid line) and $\v{S}_{\v{r}}= f(r) \left(x/r, y/r, z/r\right)$ in 3D (thick solid line) where $f(r)=1-\delta_{\v{r},0}$ (grey arrows in inset) in the $13\times 13$($\times 13$) lattice. Inset: Schematic illustration of the zero energy state when $t/J=0.1$. Local average spin of zero energy state (solid  arrows), $\langle \zeta |\bm{\sigma}_{\v{r}} | \zeta\rangle$, is perpendicular to $\v{S}_{\v{r}}$ in the main. The red arrow and blue arrow are the up spin vector and the down spin vector, respectively. Adjusted magnetization amplitudes $|\langle \zeta | {\bm \sigma}_{\v r} | \zeta \rangle |^{1/4}$ is used in the figure for clarity.}
\label{fig3}\end{figure}

An illustration of the zero-energy solution around the anti-vortex spin configuration (gray arrows) is shown in Fig. \ref{fig3}. We choose the spin texture, $\v{S}_{\v{r}}= f(r) \left(y/r, x/r,0 \right)$ in 2D and $\v{S}_{\v{r}}= f(r) \left(x/r, y/r,z/r \right)$ in 3D, where $f(r)=1$ everywhere except at the origin, $f(\v 0 ) = 0$. Since $| \v{S}_0 \rangle$ is not fixed, the spin orientation of the zero energy state at the node is assigned $| \v{S}_0 \rangle=\left(1,0 \right)^T$. Exact numerical result of the lattice model with this profile $\v S_{\v r}$, on a lattice of odd linear dimension with open boundary condition, matches our analytic solution in Eq. (\ref{eq:zeta-1}) for the same $\v S_{\v r}$ almost perfectly, $|\langle \zeta_{\mathrm{exact}}|\zeta\rangle| \approx 1$, in the small $t/J$ region.

Existence of two zero-energy states per defect is guaranteed by symmetry of the anti-vortex or the hedgehog profile which obeys $\v S_{-\v r} = -\v S_{\v r}$. This leads to an additional, inversion symmetry of the Hamiltonian under the operation $\cal I$ that moves each $\v r$ to $-\v r$. The operation gives

\ba {\cal{I}} H_K  = H_K {\cal{I}}, ~~ {\cal{I}} H_J = -H_J {\cal{I}}. \ea
The composite operator $\Theta {\cal{I}}$, which commutes with the Hamiltonian, ensures a pair of Kramers-like degenerate states. Combined with the particle-hole symmetry which is also present in the model and assuming that the calculation is done on a lattice of odd linear dimension (therefore of odd total sites), we are forced to conclude that there must exist a pair of states at zero energy which precisely corresponds to the localized solutions discussed in this section.

\begin{figure}[tb]
\includegraphics[width=0.48\textwidth]{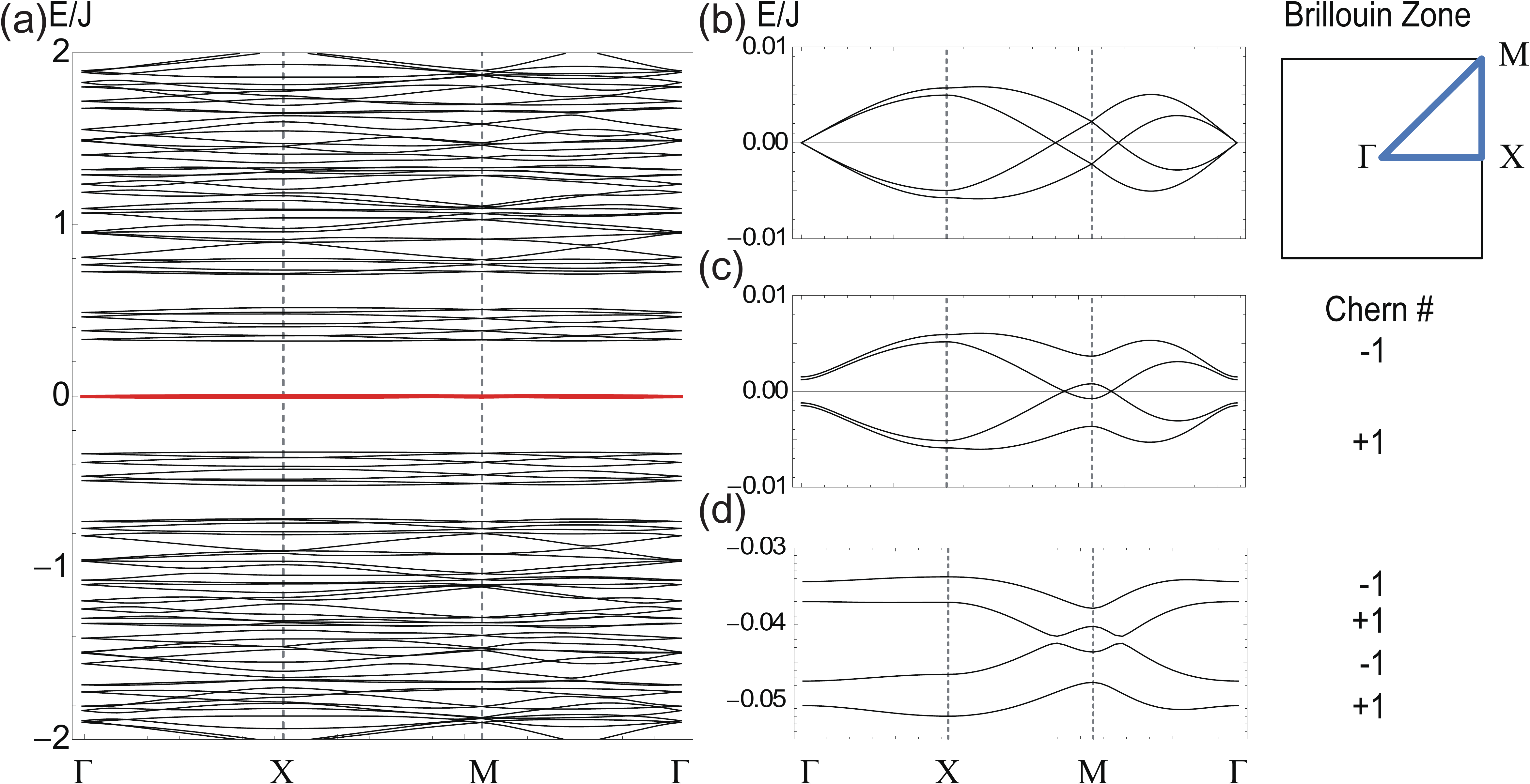}
\caption{(color online) (a) Electronic band structure for 2D M$\overline{\rm M}$ lattice,
using $t/J=0.3$  and $R=8$. Nodal bands are indicated as almost overlapping red lines. (b) Enlarged nodal bands at the zero magnetic field. (c) Magnetic field $m_z=0.05J$ is added. Outer two bands are split away and carry Chern numbers. (d) NNN hopping $t'=0.1t$ is added. All four nodal bands become non-degenerate and carry distinct Chern numbers indicated on right.}\label{fig5}\end{figure}

\section{Band structure of nodal topological lattice}
\label{sec5}

\begin{figure}[tb]
\includegraphics[width=0.45\textwidth]{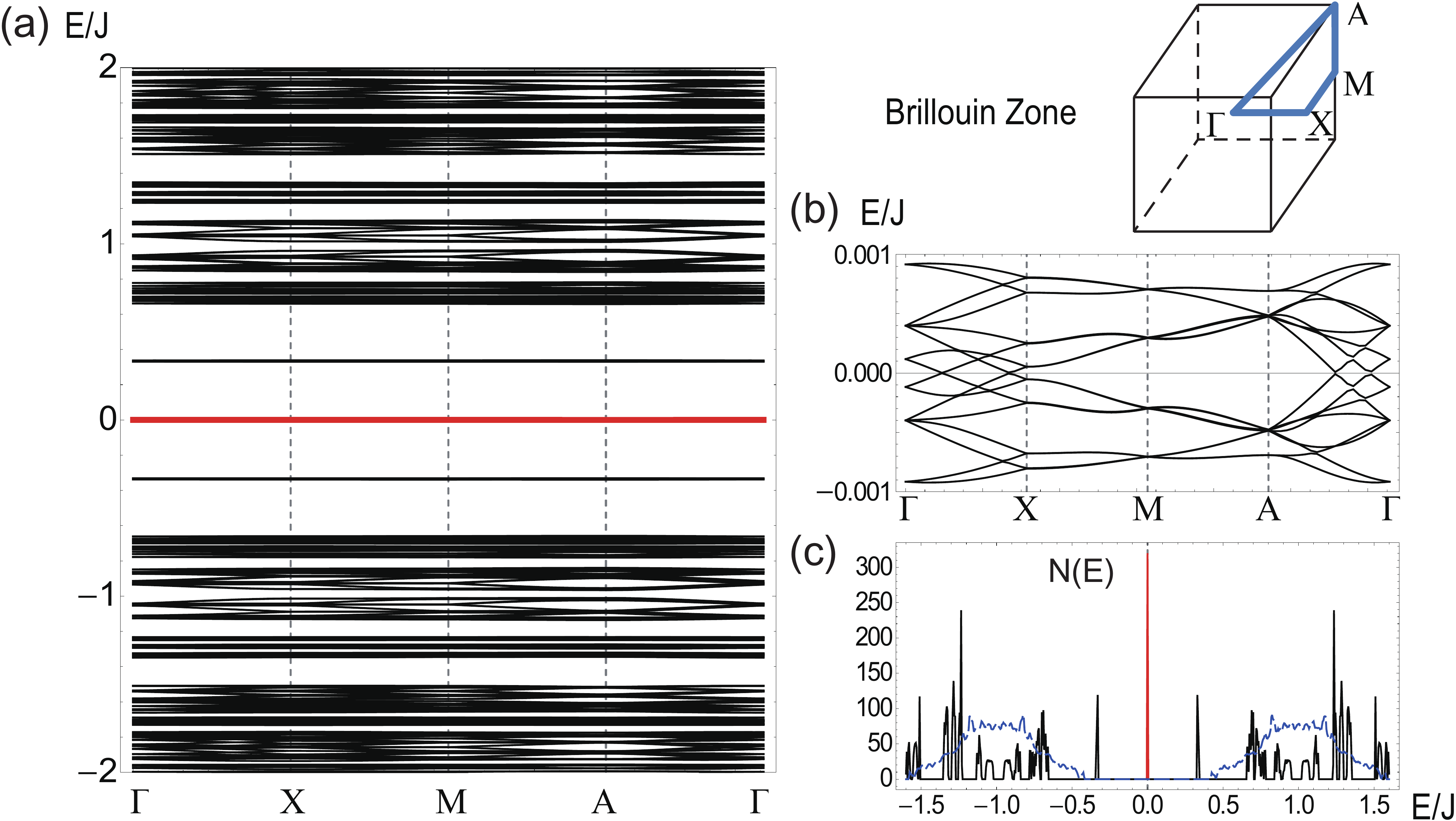}
\caption{(color online) (a) Electronic band structure for 3D simple cubic H$\overline{\rm H}$ lattice using $t/J=0.1$  and $R=8$. Nodal bands are indicated as almost overlapping red lines. (b) Enlarged view of 16 nodal bands near zero energy. (c) Electronic DOS for H$\overline{\rm H}$ lattice (black solid line, arbitrary units). A peak due to zero energy bands is clearly shown in solid red.  Zero-energy DOS disappears when the magnetic texture is a simple spiral $\v{S}=(0, \cos Kx, \sin Kx)$ (dahsed blue curve).}\label{fig6}\end{figure}

Full electronic structure coupled to the nodal lattice is worked out in Figs. \ref{fig5} and \ref{fig6}, for two and three dimensions, respectively. Nodal bands (induced by the periodic nodes in the magnetization) around $E=0$ are indicated as red curves. At the $\Gamma$ point, there are four degenerated states caused by the $\Lambda$ and $\Omega$ symmetry. Note that $\Lambda^2=-e^{i \v{k} \cdot 2 \bm{\pi}_d}$ for a specific Bloch momentum $\v k$. Qusi-Kramers degeneracy is guaranteed at the $\Gamma$ and the $M$ point (corner of the Brillouin zone) due to $\Lambda^2=-1$.

Chern numbers in 2D appear for the central bands upon breaking of time reversal symmetry. The non-trivial Chern number is the direct evidence of the topological Hall effect~\cite{NagaosaTokura}. The quasi-time-reversal symmetry $\Lambda$ is broken, for instance, by the Zeeman field $H_Z =-m_z \sum_\v{r} \bigl( c^\dagger_{\v{r},\uparrow}c_{\v{r},\uparrow}-c^\dagger_{\v{r},\downarrow}c_{\v{r},\downarrow} \bigr)$. Two outermost bands separate away from the two inners ones and take on Chern numbers $\pm 1$
as shown in Fig. \ref{fig5} (b). The $\Omega$ symmetry is broken by the next nearest neighbor (NNN) hopping, $H_{\text{NNN}}=-t' \sum_{ \langle \langle \v{r},\v{r}' \rangle \rangle} \bigl( c^\dagger_{\v{r},\uparrow}c_{\v{r}',\uparrow}+c^\dagger_{\v{r},\downarrow}c_{\v{r}',\downarrow} \bigr)$. Now all the zero-energy bands have the Chern numbers shown in Fig. \ref{fig5} (c).

The 3D simple cubic H$\overline{\rm H}$ band structure includes the 16 zero energy bands shown in Fig. \ref{fig6}. At the $M$ point, the bands are degenerate as the Kramers pair. Due to the complexity of the calculation we have not attempted the calculation of the cross-sectional Chern numbers after all the band degeneracies are split. A reasonable expectation is that both the zero-energy bands and the bulk bands carry Chern numbers for a general cross section. A better and more direct manifestation of the zero-energy states may be the enhanced density of states (DOS) around $E=0$, shown in Fig. \ref{fig6} (c).
\\

\section{Discussion}
\label{sec6}

Several recent electronic structure calculations of MnGe reveal an extremely complex band pattern of this material~\cite{Rosler,Brickson,Mirebeau2}. A rough consensus among the calculations points to the Hund coupling  $J_{\text{MnGe}} \simeq 2$ eV, much bigger than the typical bandwidth of an individual band $t \simeq \text{a few 100 meV}$~\cite{Rosler,Brickson,Mirebeau2}. The large $J/t$ assumption used throughout this work is reasonably well justified for the majority-minority band pair in MnGe, suggesting that the formation of solitonic narrow bands half way between them is likely to accompany the H$\overline{\rm H}$ phase. Extra states induced by the nodal lattice can be picked up by the DOS studies employing surface-sensitive STM or tunneling techniques. We point out that the M$\overline{\rm M}$ lattice phase has been predicted to occur in the interface region between $\mathrm{SrTiO}_3$ and $\mathrm{LaAlO}_3$~\cite{Balents}. Furthermore, the non-collinear magnetic texture attached to the $s$-wave supercoductor could provide a higher-dimensional platform for realizing Majorana states~\cite{Yazdani,Yazdani1}. Possible existence of Majorana states in nodal topological lattice in dimensions two and three will be examined in the future.

In conclusion, we addressed the electronic structure of the nodal topological lattice that have potential relevance to MnGe crystal or the $\mathrm{SrTiO}_3 /\mathrm{LaAlO}_3$ interface. We find two zero energy states trapped around each node. Wave functions for the localized states are worked out in lattice. The narrow bands formed by the zero-energy states carry Chern numbers and lead to enhanced density of states in an otherwise mid-gap region. Zero-energy modes accompanying the two- and three-dimensional non-coplanar spin structure should spur a closer examination at realization of higher-dimensional Majorana states.

\acknowledgments{We acknowledge helpful discussions with Leon Balents, Naoya Kanazawa, Bongjae Kim, Pajin Kim, Junhyun Lee, Ye-Hua Liu, Naoto Nagaosa, and Adam Nahum. J. H. H. would like to thank members of the MIT condensed matter theory group for their hospitality during his sabbatical leave. This work is supported by the NRF grant (No. 2013R1A2A1A01006430).}


\begin{thebibliography}{99}
\bibitem{comment} Localized nature of the eigenstate can be checked through calculation of the state's inverse participation ratio, for example. Inclusion of higher harmonics in the magnetization profile produces the same number of nodal states at zero energy. The existence of node-localized states appear to be independent of the exact profile of the magnetization, only requiring the presence of nodes in it.
\end{thebibliography}
\end{document}